\documentclass[journal]{IEEEtran}
\usepackage{cite}
  \usepackage[pdftex]{graphicx}
\usepackage[caption=false,font=footnotesize]{subfig}
\usepackage{txfonts}  
\usepackage{stfloats} \fnbelowfloat
\usepackage{url}
\begin{document}
\title{Increasing Launch Efficiency with the PEGASUS Launcher}
\author{\IEEEauthorblockN{S.\,Hundertmark, G.\,Vincent, D.\,Simicic,
M.\,Schneider\\}
\IEEEauthorblockA{French-German Research Institute,
Saint Louis, France\\
}
}

\maketitle
\begin{abstract}
In the real world application of railguns, the launch efficiency is one of the most
important parameters. This efficiency directly relates to the capacity of the electrical
energy storage that is needed for the launch. In this study, the rail/armature contact
behavior for two different armature technologies was compared. To this end, experiments
using aluminum c-shaped armature and copper brush armature type projectiles were performed
under same initial conditions. The c-shaped armature type showed a superior behavior 
with respect to electrical contact to the rails and in acceleration. A 300\,g
projectile with a c-shaped armature reached a velocity of 3100\,m/s and an overall launch 
efficiency including the power supply of 41\%. This is to be compared to 2500\,m/s 
and 23\% for the launching of a projectile using a brush armature. 
\end{abstract}
\section{Introduction}
Large caliber railguns are an attractive solution for long range
shipboard artillery. They allow large muzzle energies and ranges far in
excess of the capability of current deck guns \cite{mcnab, shipboard}.
For a future electrical warship, electric weapons like the
railgun and/or the high-energy laser are the logical choice, with
respect to capability, integration and economy. A large caliber
railgun is a gigawatt launcher with rail currents of several megaamperes.
Such an electrical machine requires power close to an order of magnitude above the 
power generation capability being installed ind current vessels. Ships with the 
largest installed electrical power are large, modern cruise ships. They are combining several
generators to feed the electrical drives and all the other loads. The largest vessels have a 
total power rating  
close to or above one-hundred megawatts \cite{oasis, queen_mary_2}. In the military domain,
the US-Navy recently commissioned the first Zumwalt class destroyer, with 78\,MW of installed 
electrical power \cite{zumwalt}.  
To adopt the required power level for the railgun to the generator power level 
an intermediate energy storage system is
required. Such a system allows a slow charging with lower power and a rapid discharging
during firing of the gun. 
The size of this system is correlated to the muzzle energy
divided by the overall efficiency of the launcher system. Even so modern frigates or
destroyers are rather large vessels, volume and mass carrying
capability are still limited. To maximize the launcher efficiency it is
important to reduce electrical losses. For the launcher itself,
the most important part is the armature and its contact to the rail
surface. The energy losses at this high-speed sliding contact
determine to a large extent the ability of the accelerator to convert
the energy provided to the launcher into kinetic energy of the
payload. Reduced losses at the armature/rail interface will have a
beneficial impact on the rail wear and barrel lifetime.
In a series of experiment performed at the French-German Research Institute with the
PEGASUS railgun installation two different armature types were directly compared. 
The results of this study clearly favor the widely used c-shaped aluminum armature 
type.  
\section{PEGASUS Railgun Installation}
\begin{figure}[tb!]
\centering
\includegraphics[width=3.5in]{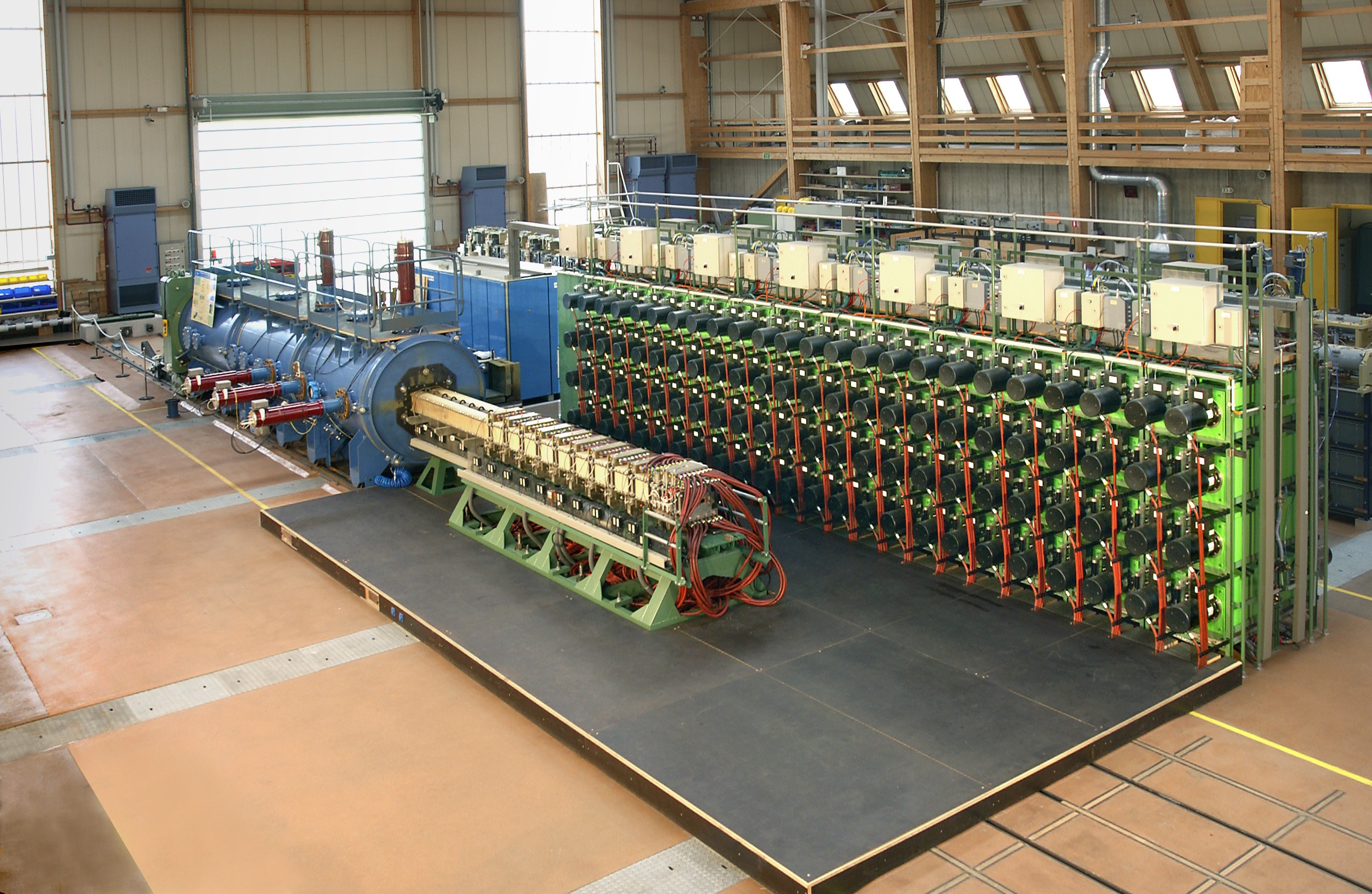}
\caption{PEGASUS railgun installation.}
\label{peg}
\end{figure}
The PEGASUS railgun installation consists out of a railgun barrel, a
10\,MJ capacitor based power supply, a 7\,m catch tank, a 50\,kW capacitor
charger and a Faraday cage with the data acquisition and experimental
control. Apart from the Faraday cage, the experimental facility is
shown in fig. \ref{peg}. The barrel used in this investigation is a 6\,m
long, 40\,mm square caliber, closed tube with twelve current injection points distributed along the
first 3.75\,m of acceleration length. This DES ({\bf D}istributed
{\bf E}nergy {\bf S}upply) scheme allows to follow the armature with
the current injection while it propagates down the acceleration
length. B-dot sensors installed along the barrel register the passage
of the armature. These signals are used to
trigger the release of portions of the energy to the different current
injection points and to calculate the velocity of the armature. The modularity 
of the power supply allows to create
a current pulse with a nearly flat plateau over much of the
acceleration time period, thus allowing for a nearly constant
acceleration. As the armature moves down the barrel, the
current from current injection points which had been triggered earlier
starts to decay and the energy being stored in the magnetic field of
this section is used to drive the current through the rest of the
barrel. Therefore, in comparison to a breech-fed barrel of the same
length, the amount of magnetic energy being stored in the barrel at
armature exit is strongly reduced, resulting in a higher launch
efficiency \cite{hun_jee}.
Just before the projectiles leaves the muzzle, an x-ray window in the barrel 
allows to take a picture of the armature. The catch tank is equipped with different
diagnostic devices, most notably 5 flash x-ray tubes, mounts for up to two high
speed cameras and several laser barriers. In addition Doppler radar systems
installed inside the tank can be used to determine the projectile velocity. 
These systems are used as an independent means to assess the velocity of the
accelerated body. The power supply is composed of 200 capacitor
modules with 50\,kJ energy capacity, each. The electric circuit
attached to the individual capacitors facilitates
the transfer of the energy to the gun. Its topology is a typical step-down converter
using a thyristor as switch and a crow-bar diode for decoupling of the
capacitor. A 28\,$\mu$H coil serves to shape the current pulse from
the capacitor. Each module is connected by a 10\,m long coaxial cable
to the launcher. The 200 capacitor modules are grouped togehter into banks of 16
modules. During a launch these banks are fired in a  sequence that is determined by the 
passage of the armature through the barrel. The rails installed in the gun barrel limit the maximum 
current to approx. 2\,MA.
\section{Armature Types}
\begin{figure}[bt!]
\centering
\includegraphics[width=2in]{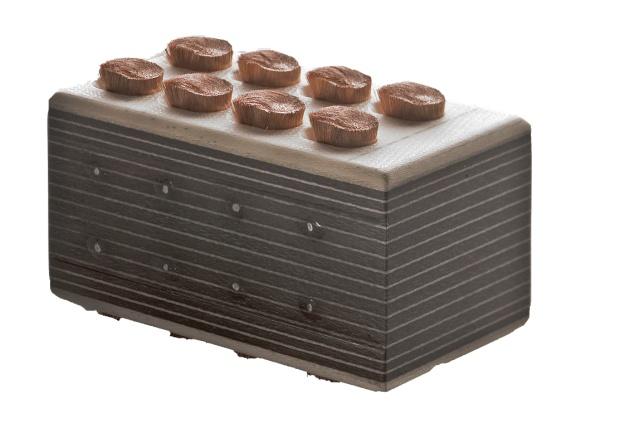}
\caption{Armature with copper brushes embedded in a reinforced plastic
body.}
\label{arm_brush}
\end{figure}

\begin{figure}[bt!]
\centering
\includegraphics[width=2in]{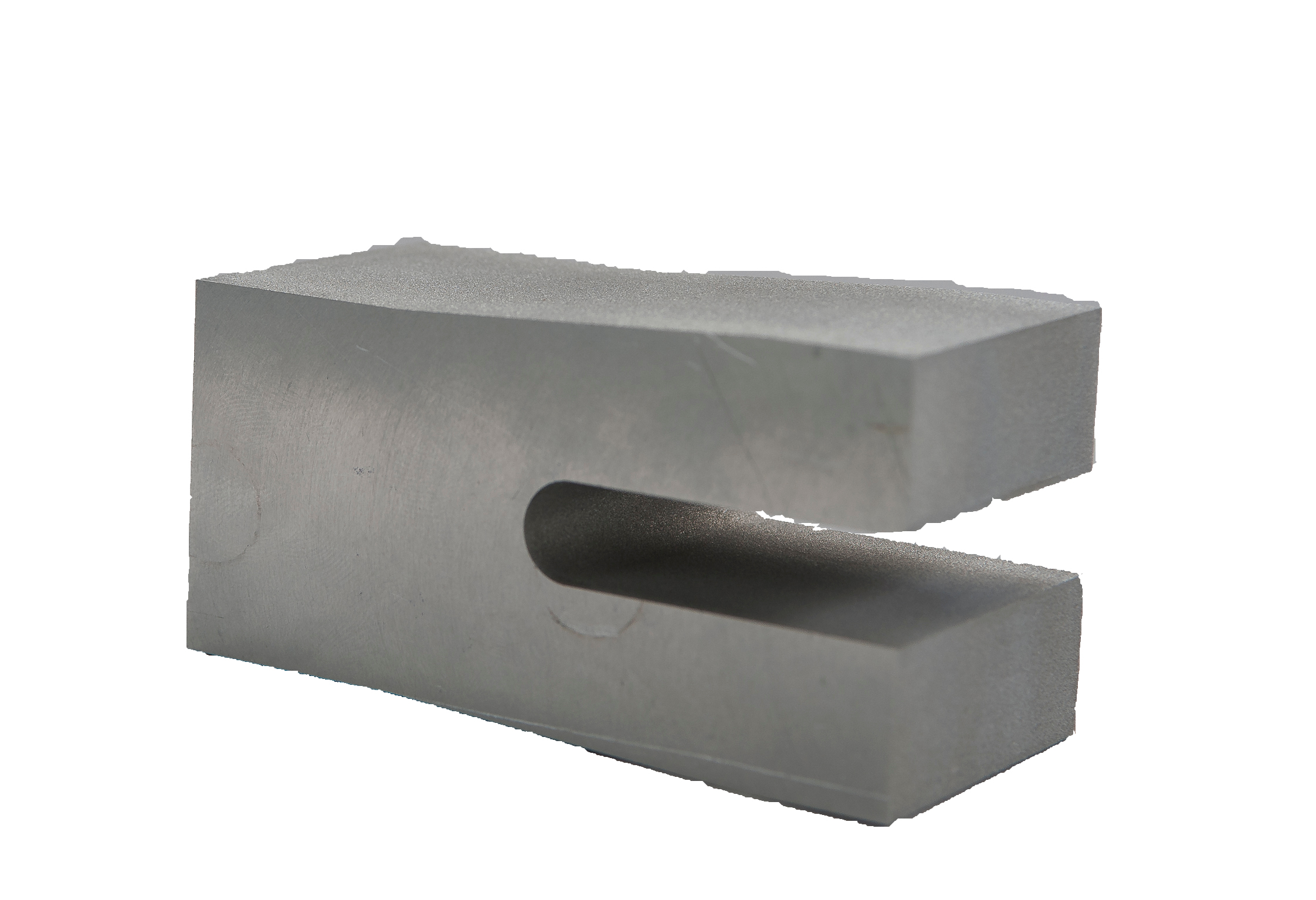}
\caption{C-shaped aluminum armature.}
\label{arm_c}
\end{figure}

The armature is the key element of the railgun. This sliding short
circuit converts electrical into mechanical energy. It has to perform under
severe and coupled mechanical, electrothermal and magnetic
constraints. Different approaches were taken to find a workable
solution that ensures loss-less electrical contact over the full acceleration period, 
while at the same minimizing the armature mass \cite{yellow}. Despite of many efforts, 
it is to be stated, that up-to-today no fully optimal armature technology has yet emerged.
One promising approach to solving the armature problem is the metal
brush armature, a concept that is extensively investigated at ISL
\cite{old_isl,payload, alu_booster}. A representative, simple fiber
brush armature is shown in \mbox{fig. \ref{arm_brush}}. Four pairs of brushes
are arranged in an isolating body with an approx. length of 7.5\,cm.
The body material of the armature shown in this figure is a sandwich made out of 
layers of glass-fiber and carbon-fiber reinforced plastic (GRP/CRP). The total mass
of this armature is 260\,g, whereof approx. 50\,g is contributed by the copper
fiber brushes. To allow for a good initial electrical contact, the brushes are produced
and mounted with 
several millimeters "overlength". This reservoir is also used to
compensate for material erosion caused by the sliding along the rail surface. A different,
simpler approach to solving the high-speed, high-current sliding
contact problem is the monolithic c-shaped armature. This type of armature is and was used
in many different experiments performed all over the world \cite{cem, watt, china, solid}.
An implementation of such an armature as being used at ISL is shown in fig. \ref{arm_c}. 
It is made out of an
aluminum alloy and has a mass of 300\,g and a length of 80\,mm. To
ensure sufficient contact pressure at the rail-armature interface the
height of the armature at the rear end of the armature arms exceeds the
nominal caliber by 2\,mm.
\section{Experiments}
\begin{figure}[tb!]
\centering
\includegraphics[width=3.2in]{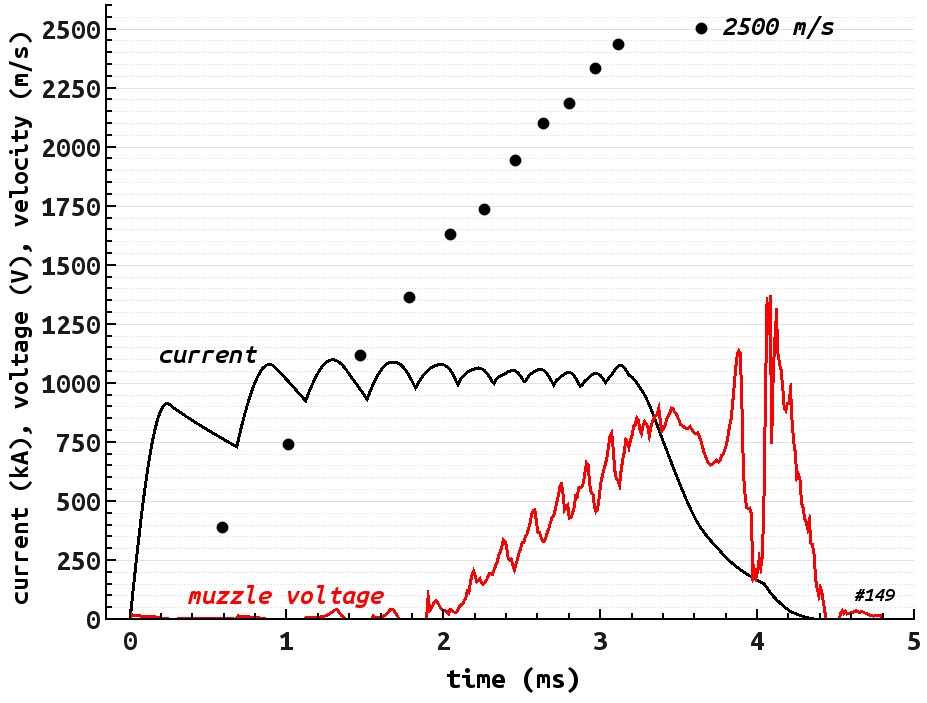}
\caption{Muzzle voltage, current and velocity trace for the launch of the brush
armature.}
\label{t149}
\end{figure}

\begin{figure}[tb!]
\centering
\includegraphics[width=3.2in]{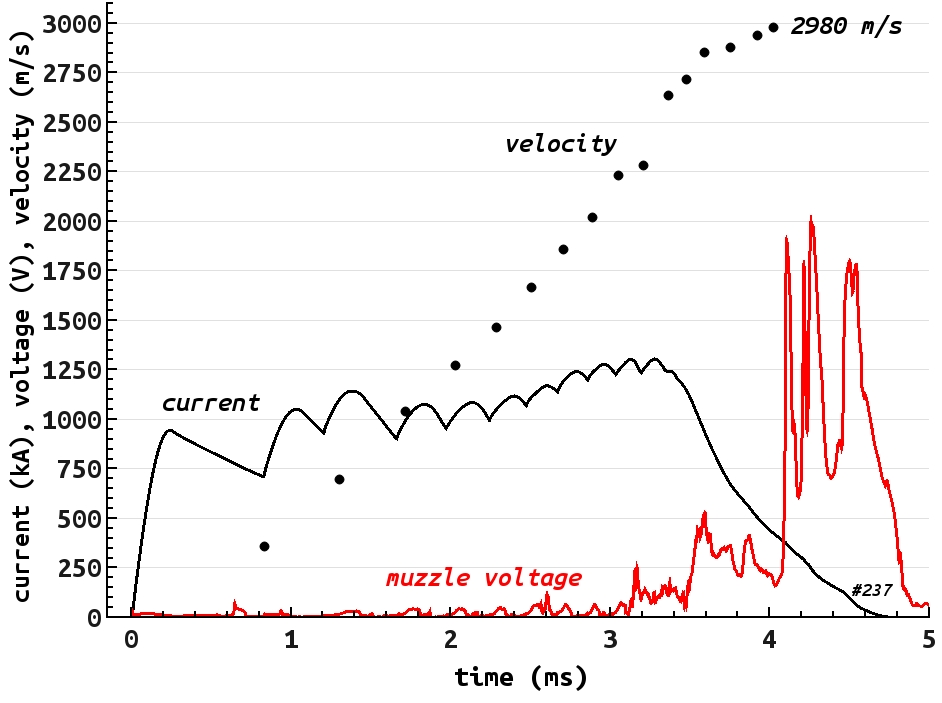}
\caption{Muzzle voltage, current and velocity trace for the launch of
the monolithic armature.}
\label{t237}
\end{figure}

\begin{figure}[tb!]
\centering
\includegraphics[width=3.2in]{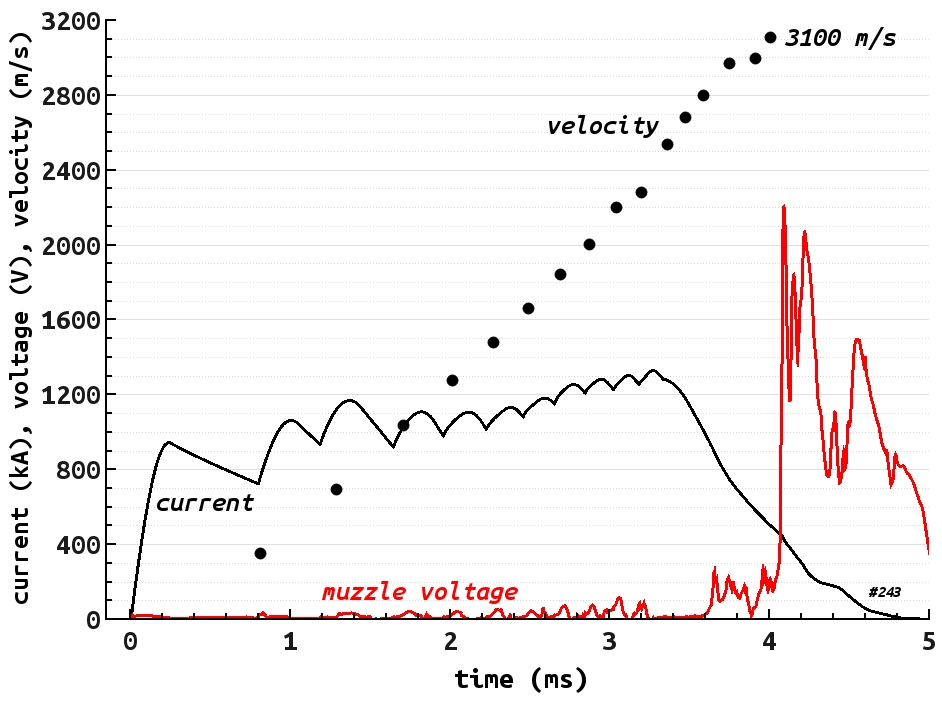}
\caption{Muzzle voltage, current and velocity trace for the second
launch of the monolithic
armature.}
\label{t243}
\end{figure}

To compare the performance of the two different armature types both were launched at the
same energy of 3.6\,MJ being stored in the capacitor modules. Results from the launch of
the brush armature is shown in fig. \ref{t149}. The current trace
shows the typical behavior of the DES structure of the PEGASUS setup.
The different banks of the power supply are triggered by the passage
of the armature of the corresponding current injection point of the
barrel. Each "fresh" injection creates a small sub-peak on top of the
DC current. Overall, in this launch, the current trace is
very close to the preferred rectangular shape, resulting in a constant
acceleration of the armature over close to the full barrel length. The current reaches
a value of approx. 1--1.1\,MA. The contact performance
can be interfered from the muzzle voltage measurement, as this value
is the voltage drop across the armature/rail interface
resistance and the armature resistance. In this shot, the muzzle voltage
value is low until approx. 2\,ms. Up to this time, the contact is a
good metal-to-metal contact and the magnitude of its resistance is
well below 0.01\,m$\Omega$.
After this time, the direct contact between the brush and the rails fail on
at least one side of the armature. The short distance in between rail
surface and brush tip is bridged by small plasma arcs. The resistance
of the plasma is larger than the metal-to-metal contact and the
muzzle voltage increases up to approx. 900\,V, corresponding to about
1\,m$\Omega$. To asses the energy lost at this contact, the current
is multiplied with the muzzle voltage and integrated until shot-out of
the armature, resulting in a value of 0.83\,MJ. This is 23\% of the 
initially stored energy in the capacitors. B-dot sensors, distributed 
along the barrel are
used to determine the time of the passage of the armature. From the
known positions and the measured passing times, the velocity of the
armature is calculated. As the current is
very close to being constant, the acceleration is constant, too. This
is reflected in the linear slope of the measured velocity curve. The
armature reaches a muzzle velocity of 2.5\,km/s and thus an
efficiency\footnote{Efficiency is here defined as energy stored in the
capacitors before the shot divided by the kinetic energy.}
of 23\%. After having discussed the launch using a brush equipped armature, a 
monolithic c-shaped aluminum armature launch is presented in the following. 
Figure \ref{t237} shows the result: For the first shot using an aluminum armature, 
the current trace shows nearly exactly the same values as in the previous discussed shot 
until approx. 2.5\,ms,
from this time on, the current begins to rise slowly up to 1.3\,MA.
This increase in current corresponds to the much lower values in muzzle
voltage during the later phase of the acceleration. The contact in
between the armature and rail surface has a lower resistance as compared
to the above described launch of the brush equipped armature and only
0.25\,MJ or 7\% of the initial energy are lost at the contact resistance. 
The muzzle velocity of the
armature was determined to be 2980\,m/s, corresponding to an efficiency of 37\%.
This is a significant improvement in efficiency to the launch under similar conditions
with the brush armature. To confirm this result, the exceptional good contact despite of the
usage of worn rails and the excellent overall efficiency, a second shot
using an aluminum c-shape armature
with similar parameters was performed. In between the first and second
shot with this type of armature, the rails of the PEGASUS barrel were
replaced and therefore the rail surface conditions for the second shot were
better. The results are shown in figure \ref{t243}. Interpreting the
muzzle voltage allows to state, that the rail/armature contact was
excellent until 3.6\,ms, only to deteriorate slightly after this time
until shot-out at 4.1\,ms. The energy lost at the contact resistance can be
determined to be 120\,kJ, less than half of the value of the similar
shot described before. With a velocity of 3100\,m/s, a launch
efficiency of 41\% is reached. The parameters and results for the
different shots are summarized in table \ref{tab_1}. Without doubt, the c-shaped aluminum 
armature in direct comparison to a brush equipped armature, is able to better convert
electrical energy into kinetic energy. Despite the large velocity of approx. 3\,km/s and
the condition of the rails, for both launches the rails were already used from other
shots, the monolithic armature performs with excellent contact behavior over the full
acceleration length.  

\begin{table}
\centering
\begin{tabular}[tbh]{|l|r|r|r|}
\hline
& brush & c-shape \#1 & c-shape \#2 \\
\hline
E$_{\mbox{cap}}$ & 3.6\,MJ & 3.6\,MJ & 3.6\,MJ\\
\hline
mass & 260\,g & 300\,g & 308\,g\\
\hline
v & 2500\,m/s  & 2980\,m/s & 3100\,m/s\\
\hline
E$_{\mbox{kin}}$ & 0.81\,MJ& 1.3\,MJ & 1.48\,MJ \\
\hline
E$_{\mbox{loss}}$ &0.83\,MJ & 0.25\,MJ & 0.12\,MJ \\
\hline
$\eta = \mbox{E}_{\mbox{kin}}/\mbox{E}_{\mbox{cap}}$&23\% & 37\% & 41\%\\
\hline
\end{tabular}
\vspace{3ex}\caption{Parameters for the comparison of brush equipped and
c-shape armature launches.}
\label{tab_1}
\end{table}


\section{Simulations}
To gain further insight a SPICE simulation was performed for 
one of the experiments with the aluminum c-shaped armature. This allows to further
investigate the launch efficiency and importance of the different energy loss processes.
In addition, with the electrical circuit being implemented into
the simulator, detailed information about values that are not easily accessible in the
experiment become available. An important example are the identification of the 
different contributions to power loss.

\subsection{Simulating the PEGASUS Railgun}
\begin{figure}[tb!]
\centering
\includegraphics[width=2in]{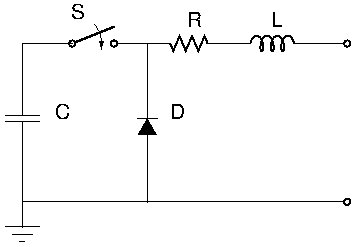}
\caption{Simplified power supply circuit.}
\label{50kj}
\end{figure}

\begin{figure}[tb!]
\centering
\includegraphics[width=3.5in]{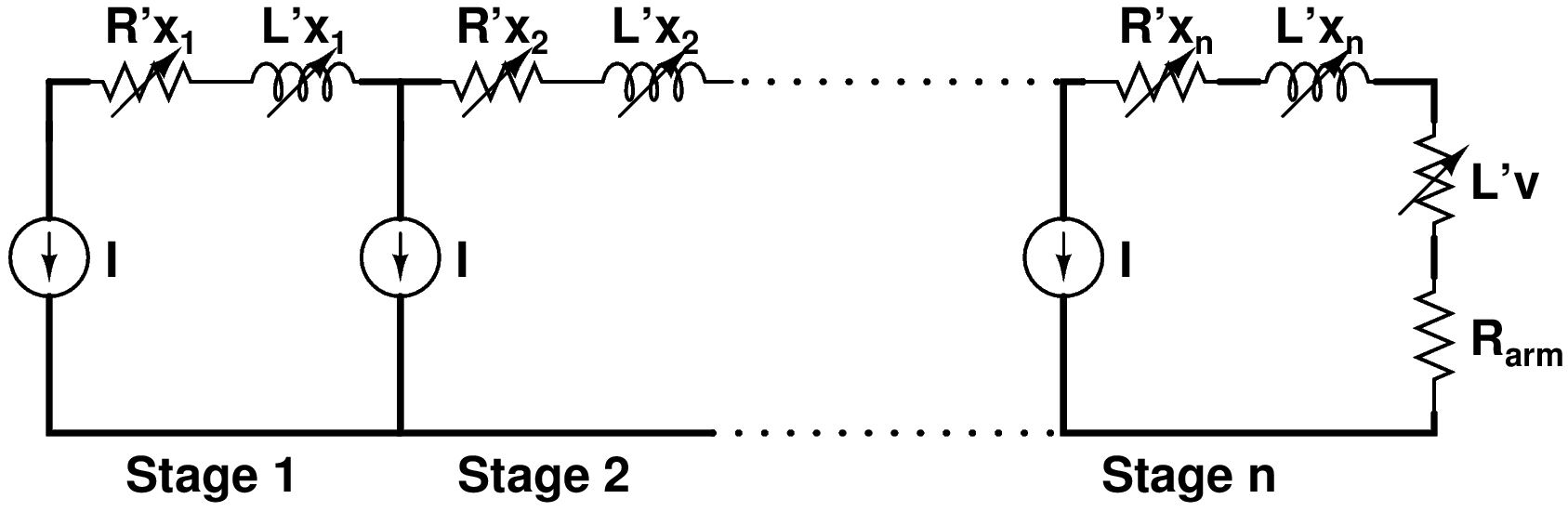}
\caption{Setup of a DES railgun composed of n stages.}
\label{des}
\end{figure}

The NGSPICE \cite{ngspice} simulator is used to simulate the electrical circuit of the 
railgun. Figure \ref{50kj} shows the circuit of the 
the power supply unit (PSU) as implemented into NGSPICE. It consists out of a capacitor, a switch, a
crow-bar diode, a resistance and an inductor. To simplify the circuit, the resistance and
inductance of the cable connecting the module to the launcher are
integrated into the resistance and the inductance of the module. In this setup 
the small difference in resistance and inductance that occurs, once
the crowbar diode becomes conducting and the switch and capacitor are
disconnected from the current flow is neglected. This  simplified circuit was
tested against experimental short circuit data and showed good overall
accuracy \cite{hun_jee}. In figure \ref{des}
the schematic representation of the PEGASUS DES railgun is depicted.  
Several capacitor modules (shown in the figure as current sources) are connected at 
different positions to the rails. The sub-circuit, including a power supply and
the rail section until the next current injection point is called a
stage. As the armature propagates through the rail section of a
stage, the inductance and resistance grows linearly with the distance traveled
up to its maximum value (determined by the length of the stage). This is reflected 
in the circuit by the variable
resistance $R'x_n$ and inductance $L'x_n$, where $x_n$ stands for the
path length the armature has traveled within the $n^{th}$ stage. 
After the armature has propagated through one stage, it enters the
next stage. Once all current injection points are passed, all the capacitor
modules are connected in parallel
via the rails. In addition to the rail resistance, the resistances L'v and
R$_{\mbox{arm}}$ are taken into account, too. It is important to note, that current
injected at different positions, "sees" different values of resistance and inductance.
Thus the DES system is a fairly complex engine, combining currents charging inductances,
while at the same time inductances from stages with dropping current inflow
convert magnetic energy into electrical current flow.

\subsection{Simulating the C-shaped Armature Launch}
\begin{figure}[tb!]
\centering
\includegraphics[width=3.5in]{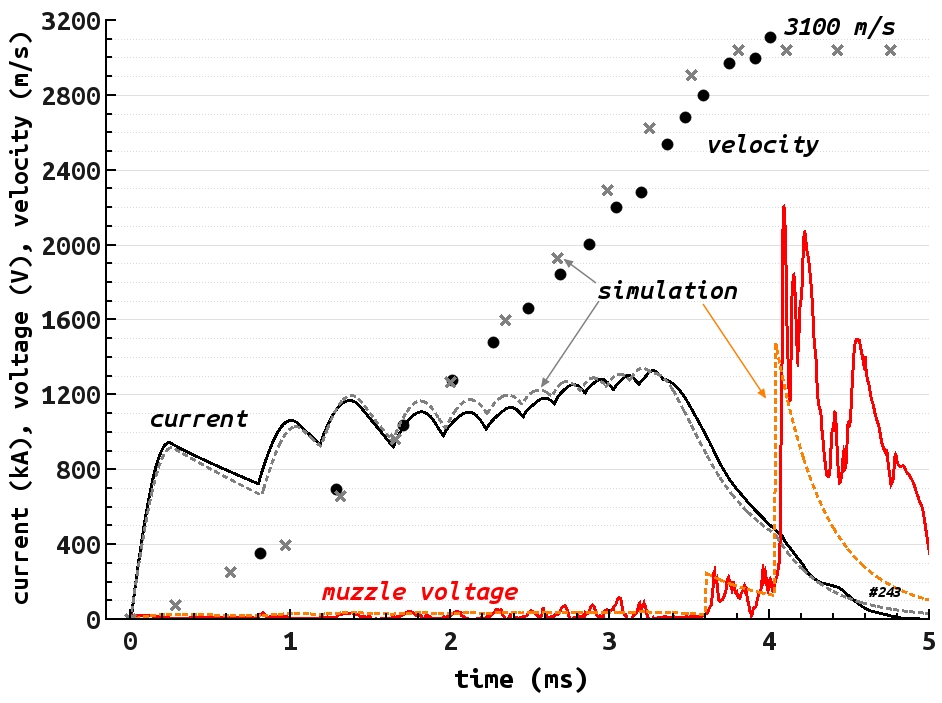}
\caption{Comparison of experiment and simulation for the c-shaped
armature launch \#2. Shown are the currents, muzzle voltages and
velocities. For the experimental data solid lines and
black dots are used, the simulation data is represented by broken lines and
tilted crosses.}
\label{t243_sim}
\end{figure}

Due to the interplay of the different stages, the DES setup is more difficult 
to simulate than the simple, breech fed railgun. 
In figure \ref{t243_sim} the results from the simulation for the second
launch is compared to the experimental data. The simulated current does
agree quite well with the measured, experimental current data. The muzzle voltage is
determined by the resistance of the armature including its contacts to the two rails.
The value of this resistance is not exactly known during sliding contact. To be able to
compare experiment and simulation, the armature and plasma
resistance (small bumb after 3.6\,ms) in the simulation was adjusted such that the energy loss at
the armature until shot-out is the same for both cases.
The acceleration and therefore the velocity is well matched between
simulation and experiment. Using this simulation, with the main launch
parameters being well described by the simulation, it is possible to
investigate the acceleration process in more detail.   
\subsection{Power of the Railgun System}
\begin{figure}[htb!]
\centering
\includegraphics[width=3.5in]{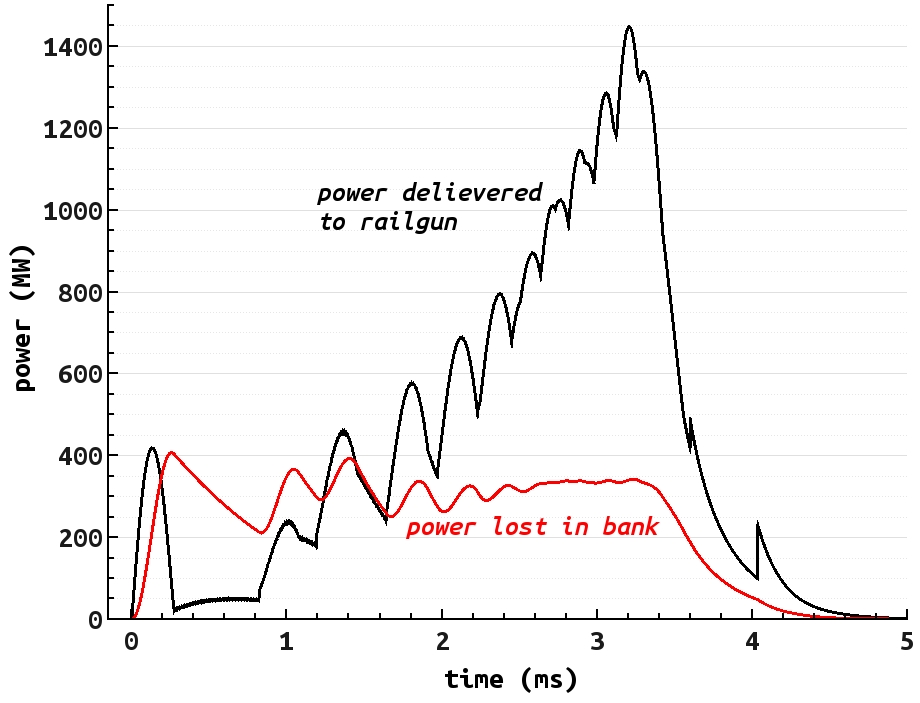}
\caption{Power delivered from the PSU to the railgun and power lost
in the PSU including the cables.}
\label{power_1}
\end{figure}

\begin{figure}[htb!]
\centering
\includegraphics[width=3.5in]{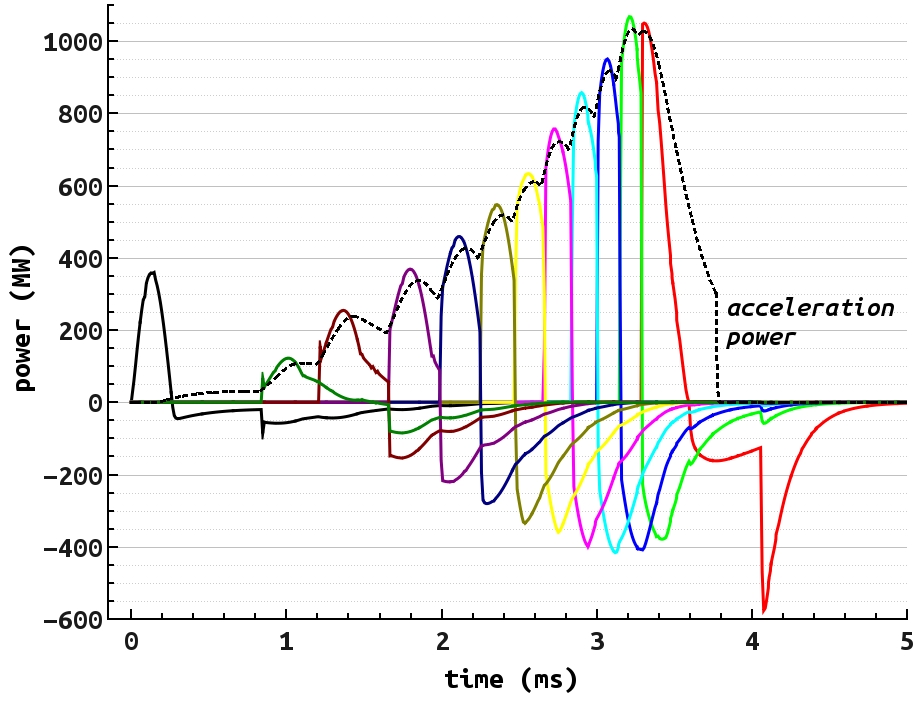}
\caption{Power required to build up the magnetic field in the rail
sections of the different stages, overlaid with the power being used to
accelerate the armature.}
\label{p_rail}
\end{figure}

In the simulation, the power being delivered from the individual banks
of the PSU to the railgun can be accessed. For every instance in time, 
the total power supplied to the gun is the sum of these contributions. 
In figure \ref{power_1} this power being delivered to the railgun is shown. 
The trace consists out of 2 parts. First we have a bump of up to 400\,MW from start to
0.3\,ms, this is followed by a rising slope until 3.1\,ms. When the experiments starts,
the cables to the railgun and the rails from the breech to the starting position of the
armature are current free. Once triggered, the current flows and the volume up to the
armature is filled with magnetic energy. During this time, the PSU has to overcome the
counter voltage of the inductance from the cables and the rail section until the armature
starting position. Only if the current has reached a large enough
value to overcome the initial friction of the armature does the acceleration process start. The current
increases further until the energy of the corresponding bank is spent. This is marked by
the downturn of the power trace up to a time of 0.3\,ms. It takes until 0.8\,ms for the
armature to reach the next injection point. From 0.3\,ms to 0.8\,ms the acceleration is
driven by the decaying magnetic field of the first stage. After the second bank has
triggered, the power delivered by the PSU to the gun describes a steeply raising function, 
reaching  a peak value of 1.45\,GW at the end of the acceleration process. The main reason
for this increasing power is the speed voltage $IL'v$ which grows linearly with the
velocity and is multiplied by the nearly constant current. Or to
phrase the same fact slightly different, the banks are current sources which must overcome 
a larger voltage with increasing
armature velocity, thus the energy in the banks is discharged in smaller time periods.
Integrating the power delivered to
the terminals of the railgun barrel over time results in the energy
being supplied to the railgun. For this shot, this energy amounts to
65\% of the initial energy stored in the capacitor banks. The remaining part is 
spent in the PSU including the cables. To investigate this, the power lost in the bank
including the cables is shown in figure \ref{power_1}, too.
Apart from the initial acceleration phase, this value is relatively
constant at 350\,MW. Integrated over time results in the remaining
35\% of the initial energy. One consequence of this behavior is, that the
relative power losses caused by the components of the PSU becomes smaller, the faster the
projectile is. This will result in an increased efficiency for shots with a higher end-velocity.
\begin{figure}[tb!]
\centering
\includegraphics[width=3.5in]{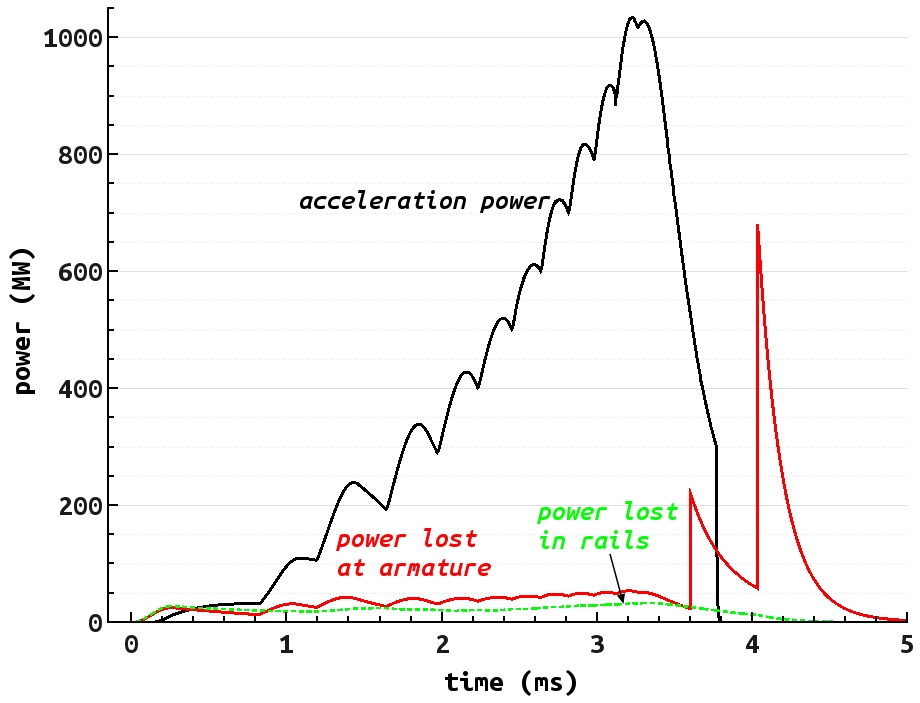}
\caption{Power used for accelerating the armature mass, power lost at
the armature and power lost in the rails of the DES railgun.}
\label{power_2}
\end{figure}
The power that is used to accelerate the armature is calculated as
time derivative of the kinetic energy of the armature. In the
simulation, the average acceleration power over the full launch period
is 375\,MW. In figure \ref{power_2} it is shown, that this power is
strongly raising with acceleration time. At shot-out, the power has
surpassed 1\,GW. This increase is easy to understand. Mechanical power
is defined as force times velocity. In this railgun launch the current is
approximately constant, so is the acceleration force. This means, that
to first order, the acceleration power grows linearly with the
velocity. Two smaller contributions to the power loss in the DES
launcher are from the voltage drop at the armature (armature
resistance and armature/rail surface contact resistance) and the
losses from the rail resistance. The DES system reduces the rail losses
drastically as compared to a breech feed system (as shown in
\cite{hun_jee}). Nevertheless a heating level of approx. 30\,MW is
not negligible with respect for rail temperature increase, especially when
considering a scenario which involves the firing of several successive rounds. 
Comparing figures \ref{power_1} and \ref{power_2} shows two things: 1) the power 
lost in the components of the banks is by a factor of approx. 10 larger than the power lost
in the rails. 2) the acceleration power, the armature loss power and rail loss power 
do not add up to
the power delivered to the railgun. The "missing" power is used
to fill the railgun stages with magnetic energy and is stored in the
inductance of the rails. In figure \ref{p_rail} the power used
to build up the magnetic field in the rail section which correspond to
the different stages is shown. It can be seen, that once the armature has
propagated into the subsequent stage, the magnetic field in the previous
stage decays and the power becomes negative. This means the magnetic
field from previous stages is used to drive the current and therefore
accelerate the armature through the following stages. The result of
this behavior is, that a
large part of the gun barrel is already "emptied" from the magnetic field
when the armature leaves the barrel at 4\,ms. From the figure one can
determine, that only stages 10 to 12 still contribute at shot-out. In an
ideal railgun, half of the power delivered to the gun goes into building 
up the magnetic field and half is converted into kinetic energy of the
short-circuit \cite{green}. This is verified by overlaying the
acceleration power onto the power needed to build up the magnetic field
in figure \ref{p_rail}. Owing to an assumed 5\% loss due to friction, 
the acceleration power is slightly smaller than the magnetic field power.
But apart from this small deviation, the equipartition between kinetic
energy and magnetic energy is clearly seen. 
Inspecting figure \ref{power_1} and figure \ref{p_rail}, there seems to be a 
contradiction: when adding up the acceleration power and the power
required to build up the magnetic field in figure \ref{p_rail}, this value is larger 
than the power being delivered to the railgun (figure \ref{power_1}). 
For example at the peak at 3.3\,ms about 1.45\,GW is being delivered to the gun, 
but the magnetic field built-up and the acceleration power add to more than 2\,GW.
This conflict of seemingly missing power is solved, when taking into account the power from the 
decaying magnetic field, which is the negative part of the curves seen in figure \ref{p_rail}.
Another detail that can be
deduced from this figure, is that for the stages 1 to 5, the energy being
available in the banks connected to these stages is not sufficient to
fully make use of the length of the stages, resulting in a strong decay of the
power level, even so the armature has not yet left the stage. 
Finally one more number should be calculated: The specific efficiency of
the railgun. Usually electric generators and engines are assigned an
efficiency which does not take into account losses from previous stages
of power conversion. This number becomes important if the quality and
further possibility of optimization has to be judged. For the PEGASUS
barrel
we can calculate an efficiency specific to this linear motor as:
\begin{equation}
\eta^{\star}=\frac{E_{{delivered}}}{E_{{kin}}}
\end{equation}
For our simulation this $\eta^{\star}$ computes to 67\%. Comparing the different
subsystems of the railgun, it can be deduced, that the launcher has a better
efficiency than the PSU including cables, which has an efficiency of 65\%.
A further improvement of the overall efficiency could therefore be
achieved by reducing the losses associated with the cables from the PSU
to the railgun.
\section{Summary}
In an experimental investigation the behavior of a copper brush
equipped armature was compared to an aluminum c-shaped armature type.
For this, 3 launches at 3.6\,MJ initial energy were performed and
current, muzzle voltage and velocity were measured. The main difference
in between the behavior of the two armature types was the amount of energy that was
lost at the rail/armature interface. For the brush armature this loss
amounted to 23\% of the initial energy, while the c-shaped aluminum
armatures had better contact over the full acceleration period 
and lost only 7\% in the first launch and 3.3\% in the second. This change in launch behavior
translated in an increase of the shot-out velocity of the projectile
from 2500\,m/s for the brush armature to 3100\,m/s for the c-shaped
armature. These velocities were reached with an overall launch efficiency, including the
power supply of 23\% (brush armature) and 37\% and 41\% for the two launches
with aluminum c-shaped armatures. The experiments performed
showed that under the given experimental conditions, the aluminum
c-shaped armatures performed much better in converting electrical energy into kinetic energy
than a brush armature of about the same mass. 
To gain further insight into the launch performance of the c-shape armature launch,
a SPICE simulation for the c-shape armature launch was carried out.
The current, muzzle voltage and velocity of the experiment could be reproduced by the
results of the simulation. Using this
simulation, insight into the power levels involved in the launch were
gained. The PSU delivers a power of up to 1.45\,GW, of this a nearly 
constant power of approx. 350\,MW is lost in resistances of the PSU 
including the cables. The acceleration power
is on average 375\,MW and peaks a little above 1\,GW. The same power
is used to build up the magnetic field inside of the DES stages.
During the launch the stages which the armature has already left
support the acceleration by reconverting the magnetic energy to drive
the current through the short-circuit. The power losses from
rail and armature resistance are only minor compared to the power used
for acceleration. One obvious result of this investigation is, that a further increase 
in launch efficiency can be accomplished by reducing the PSU and cable
losses.
\section*{Acknowledgment}
This research was supported by the French and German Ministries of Defense.

\end{document}